\documentclass[a4paper,10pt,twoside]{cpc-hepnp}

\usepackage{multicol}
      \usepackage[pdftex]{graphicx}
        \DeclareGraphicsExtensions{.pdf, .jpg}
\usepackage{booktabs}
\usepackage{amssymb,bm,mathrsfs,bbm,amscd}
\usepackage[tbtags]{amsmath}
\usepackage{lastpage}
\def\ie{{\sl i.e.}}

\def\ppLL{\ensuremath{\bar{\mathrm{p}}\mathrm{p}\to\overline{\Lambda}\Lambda}}
\def\NNb{\ensuremath{\overline{\mathrm{N}}\mathrm{N}}}
\def\RE{\mathop{\Re{\rm e}}\nolimits}
\def\IM{\mathop{\Im{\rm m}}\nolimits}
\begin{document}

\fancyhead[co]{\footnotesize J.-M. Richard et al: Constraints on spin observables}

\footnotetext[0]{Received \today}

\title{Model-independent constraints on spin observables%
\thanks{Invited talk by Jean-Marc Richard at NSTAR 2009, IHEP, Beijing, China, April 19--22, 2009}
}

\author{%
Jean-Marc Richard$^{1}$\email{jean-marc.richard@lpsc.in2p3.fr}%
\quad Xavier Artru$^{2}$%
\quad Mokhtar Elchikh$^{3}$
\quad Jacques Soffer$^{4}$
\quad Oleg Teryaev$^{5}$
}
\maketitle

\address{%
1~(LPSC, Universit\'e  Joseph Fourier, CNRS/IN2P3, INPG, Grenoble, France)\\
2~(Universit\'e de Lyon, IPNL, Universit\'e Lyon 1 and CNRS/IN2P3, Villeurbanne, France)\\
3~(Universit\'e des Sciences et de Technologie d'Oran, El Menauoer, Oran, Algeria)\\
4~(Physics Department, Temple University, Philadelphia, PA 19122 -6082, USA)\\
5~(Bogoliubov Laboratory of Theoretical Physics, JINR, 141980 Dubna,  Russia)\\  
}

\begin{abstract}
We discuss model-independent constraints on spin observables in exclusive and inclusive reactions,  with special attention to the case of photoproduction.
\end{abstract}



\begin{multicols}{2}
\section{Introduction}
The role of spin observables\cite{Bourrely:1980mr,Leader:2001gr,Ohlsen:1972xx} hardly needs to be motivated, especially in front of this audience of the NSTAR conference. The problem is to find a good compromise between spin experts, who would like to measure everything, and spin skeptics, who are reluctant to support the developments of polarized beams and targets.

The question has been often raised in the literature of the minimal set of observables that is required to reconstruct all the amplitudes of a reaction, up to an overall phase. Our approach is more empirical: given the data obtained from a first run of measurements, how much uncertainty is left for the remaining observables, and which new observable will better discriminate among different options?

In more technical words, any spin observable such as polarisation or spin transfer is typically normalised to vary in $[-1,+1]$.  However a set of $n$ observables $\{\mathcal{O}_i\}$ is often limited to a small fraction of the hypercube $[-1,+1]^n$. This means that if a few observables are already known,  any further observable is probably constrained into an interval much smaller than $[-1,+1]$. It is not necessary to measure an observable that is already much limited, except for cross checking. It is preferable to focus on observables that provide really new information.

These constraints are expressed by identities or inequalities relating various observables, which are consequence of positivity. An interesting aspect, that will not be covered her due to the lack of time, is whether the constraint is classical, or requires a quantum treatment of the spin configurations.

This contribution will be organized as follows. In Sec.~2,
we treat as a preamble the case of pion--nucleon elastic scattering, and of the strangeness-exchange reaction \ppLL, and then discuss the case of the photoproduction of pseudoscalar mesons, in the light of recent measurements. We briefly present in Sec.~3, some results dealing with inclusive reactions. A survey of the methods is given in Sec.~4. Section~5 is devoted to some conclusions.
\section{Exclusive reactions}

\rm
\label{sec:excl}
\subsection{Spin-0--spin-1/2 elastic scattering}
\begin{center}
\includegraphics[width=.65\columnwidth]{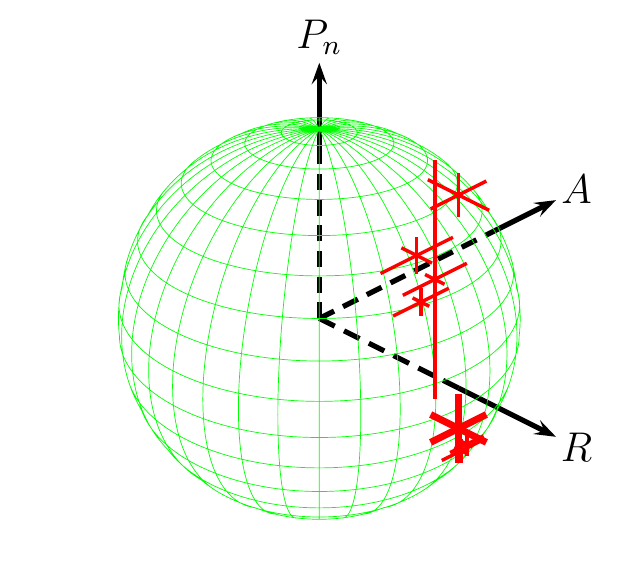}
\end{center}
\vskip -.2cm
\figcaption{\label{excl:fig:abaev}
Spin observables  for $\pi^-\mathrm{p}$ elastic scattering at 0.573 
and 0.685 GeV$/c$, compared to the unit
sphere. For the data, see \protect\cite{Artru:2008cp} and refs.\  there.
}
In the case of $\pi\mathrm{N}$ elastic scattering, there are three independent observables, which can be chosen as the polarisation $P_n$ (which coincides with the analysing power) and the spin-rotation parameters $A$ and $R$.  They are submitted to the well-known identity
\begin{equation}
P_n^2+A^2+R^2=1~.
\end{equation}
This means for instance that if $|P_n|$ is large, both $A$ and $R$ should be small.
Independent measurements of these three observables indicate that the identity is well satisfied, within error bars.
Clearly from the above identity one can derive disk constraints such a $A^2+P_n^2\le 1$ for each pair of observables. 

In other exclusive reactions, such disk constraints $\mathcal{O}_1^2+\mathcal{O}_2^2+\cdots\le1$ are often encountered.  An explanation is that the operators (of which the spin observables are the expectation values) anticommute.\cite{Artru:2008cp}\@  More exotic shapes are found in other reactions.
\end{multicols}


\subsection{\boldmath $\ppLL$\unboldmath}
This reaction has been extensively studied by the PS185 collaboration at the LEAR facility of CERN. Thanks to their weak decays, the polarisation of both $\Lambda$ and $\overline{\Lambda}$ spins can be measured.   
Interesting results came  from a first set of runs, indicating a striking correlation between the two spins in the final state. In particular, the spin-singlet state is very much suppressed as compared to the triplet.

\vglue .5cm

\begin{minipage}{.7\textwidth}
\begin{center}
\includegraphics[width=.9\textwidth]{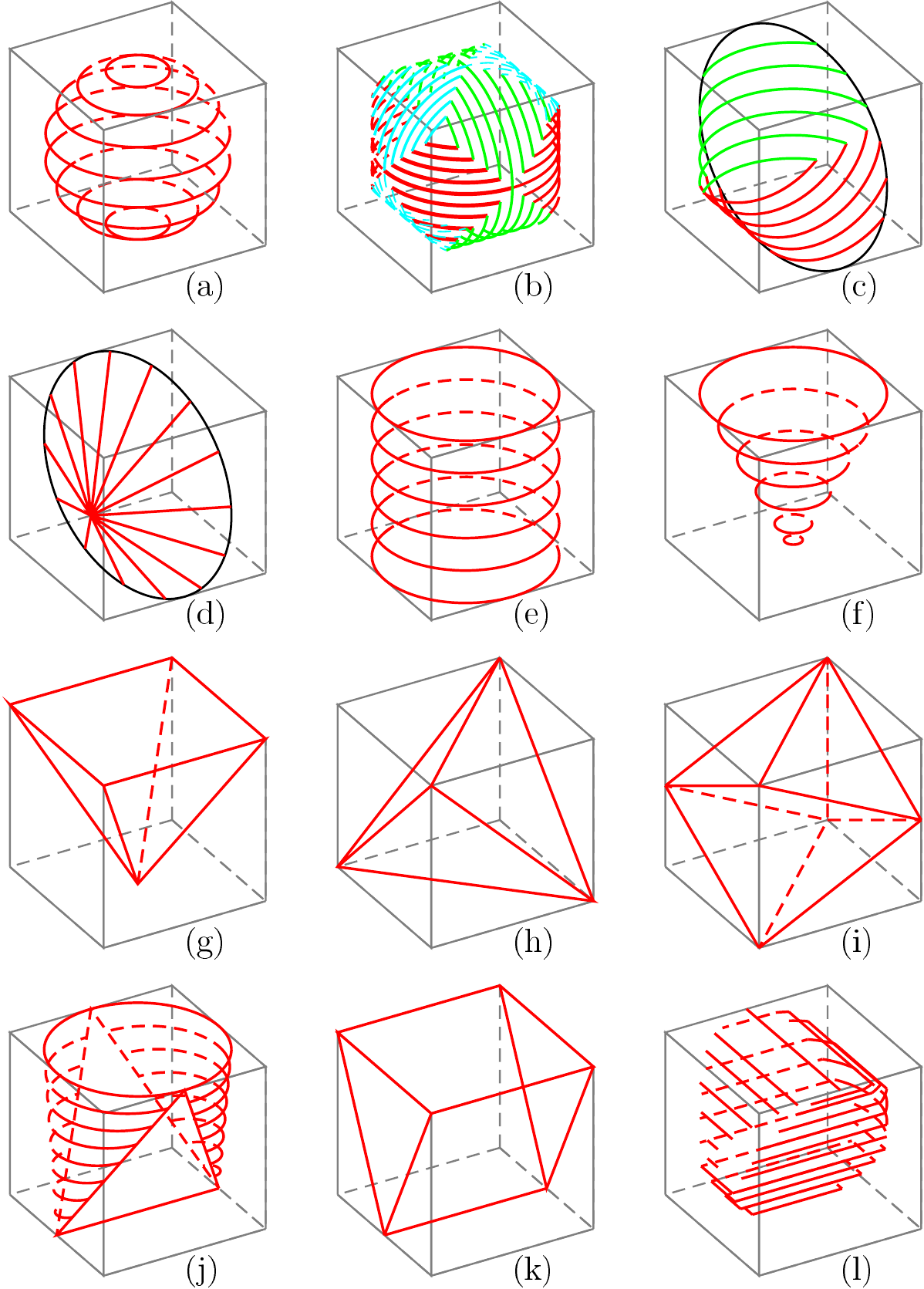}\end{center}
\end{minipage}
\begin{minipage}{.29\textwidth}
\figcaption{\label{excl:fig:lim3d}%
Some allowed domains encountered in  simulating randomly three observables: 
the unit sphere (a), 
the intersection of three orthogonal cylinders of unit radius (b),
the intersection of two cylinders (c),
  or a slightly smaller  double cone (d), 
  a cylinder (e),
  a cone (f), 
a pyramid (g), 
  a tetrahedron (h),
  an octahedron (i),
  a ``coffee filter'' (j),
an inverted tent (k), 
and the intersection of two cylinders and a dihedral (l).  For clarity, part of the limiting surface is sometimes removed.}
\end{minipage}
%

\clearpage

\begin{multicols}{2}

This motivated a number of theoretical studies. Unfortunately both models \`a la Yukawa, with $\mathrm{K}$, $\mathrm{K}^*$ exchanges and quark-based models with $q\bar{q}$ annihilation and $s\bar{s}$ creation  were able to reproduce quite well the most saillant feature of these early data: the spin-singlet fraction is compatible with zero, i.e., the reaction occurs in a spin-triplet state. It was then decided to measure the reaction with a transversely-polarised target  and to focus on $D_{nn}$ and $K_{nn}$ which measure how the transverse polarisation of the proton is modified in the $\Lambda$ or transferred to the $\overline{\Lambda}$.  It was estimated that the two above classes of models give drastically different predictions for these observables, one with $D_{nn}>0$ and another with $D_{nn}<0$. A third mechanism was also suggested, where the $s\bar{s}$ pair, instead of being created out of the vacuum, is extracted from the polarised sea of the nucleon or antinucleon.

When the data of $D_{nn}$ eventually came, it was disappointing to get an almost vanishing value, making it difficult to distinguish among models. We now realise that this $D_{nn}\simeq0$ could have been anticipated from a more careful analysis of the data obtained without target polarisation. There are in particular model-independent inequalities
\begin{equation}
D_{nn}^2+C_{mm}^2\le 1~,
\quad
D_{nn}^2+C_{ll}^2\le 1~,
\end{equation}
which indicate that at energies and angles where $|C_{mm}|$ or $|C_{ll}|$ is large, $D_{nn}$ should be small. Here, $C_{ij}$ is the spin correlation in the final state, for the longitudinal ($l$) or sideways ($m$) directions in the scattering plane.

Anyhow, the possibility of measuring many different spin observables for the \emph{same} reactions motivated further studies on the systematic of the identities and inequalities, which are summarised in \cite{Artru:2008cp}. In particular,  the domain allowed for triples of observables was considered. Among the results, one could notice
\begin{itemize}\itemsep -3pt
\item
There are cases where for three observables, none of the pairs is constrained (\ie, the whole square $[-1,+1]^2$ is allowed, but the triple is severely restricted, for instance inside a tetrahedron whose volume is only 1/3 of the cube $[-1,+1]^3$.  See Figs.~\ref{excl:fig:lim3d} and \ref{excl:fig:2-3-8-2-4-7}.
\item
Exotic shapes are found for the limiting domain (see Fig.~\ref{excl:fig:lim3d}), such as the ``coffee filter'' of Fig.~\ref{excl:fig:2-3-8-2-4-7}.
\end{itemize}
\subsection{Photoproduction}
The study has been extended to  photoproduction of mesons, such as 
\begin{equation}
\gamma + \mathrm{p}\to \Lambda + \mathrm{K}~,
\end{equation}
for which many new data have been taken recently.
%
\begin{center}
\includegraphics[width=.40\textwidth]{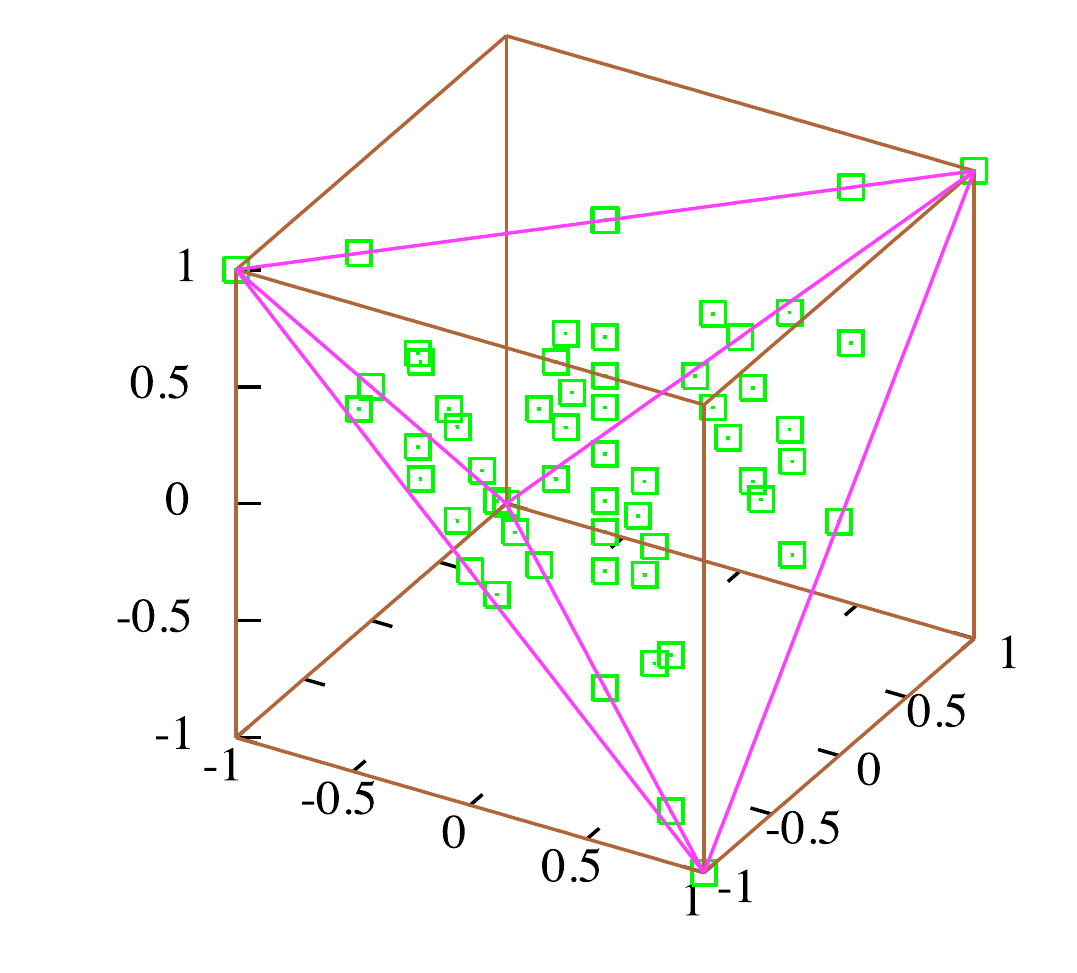}
\includegraphics[width=.40\textwidth]{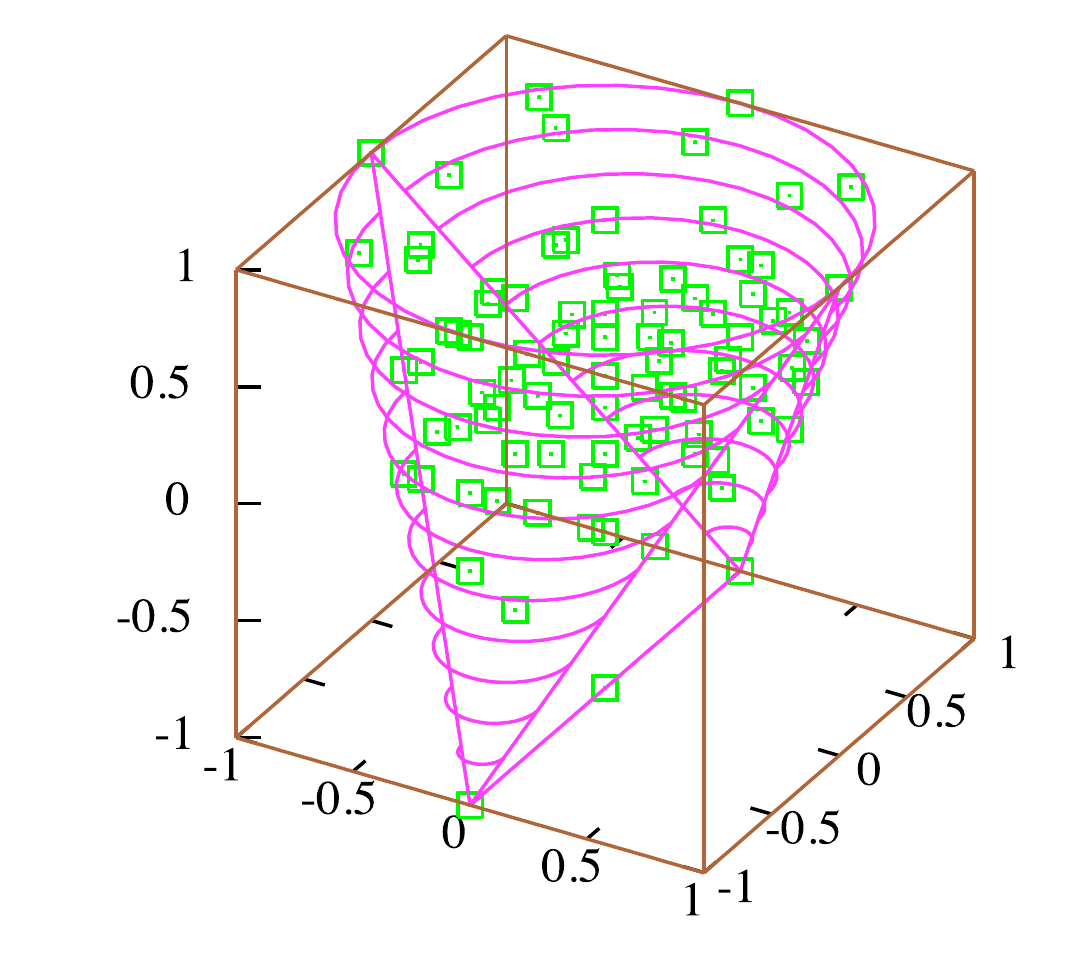}
\end{center}
\figcaption{\label{excl:fig:2-3-8-2-4-7} The domain for  $\{P_n,A_n,D_{nn}\}$ (top) 
and $\{P_n,C_{mm},C_{nn}\}$ (bottom). The dots correspond  to randomly-generated  fictitious amplitudes which are used to revel the domain before its boundary is rigorously established.}
%

There is a considerable literature on this reaction. Many identities and inequalities among observables have been written down. But the aim was mostly to determine which minimal set of observables is required  for a full reconstruction of the amplitudes, up to an overall phases.

The point of view here is slightly different: given one or two spin observables, what is the domain left for 
the other observables? The analysis indicates in particular:\cite{Artru:2006xf}
\begin{itemize}\itemsep-3pt
\item
The same limiting shapes as for $\ppLL$ are observed. In particular, for the three observables of rank 1, the analysing power $A$, the hyperon polarisation $P$ and the beam asymmetry $\Sigma$, there is the same tetrahedron constraint as above. This means that all pairs such as $\{A, P\}$ are unconstrained, but the allowed domain for the triple is only 1/3 of the cube.
\item
As for $\ppLL$, any triple of observables is correlated: if one knows two of them, any third one is constrained.
\item Some of the identities or inequalities among observables have been tested recently.\cite{Lleres:2008em}\@ It was found that the latest CLAS and GRAAL measurements are compatible.
\end{itemize}

To be more specific, let us denote $O_i$ the transfer from an oblique polarisation of the photon to the polarisation of the recoil baryon, and $C_i$ the analogue for a circular polarisation.  The index $i$ refers to the component in a frame $\{\hat{x},\hat{y},\hat{z}\}$ attached to each particle, $\hat{y}$ being normal to the scattering plane, and $\hat{z}$ along the momentum in the centre-of-mass frame. 

Examples of disk inequalities for pairs are:
\begin{equation}
\begin{aligned}
\Sigma^2+O_x^2& \le 1~,\\
\Sigma^2+O_z^2&\le 1~,\\
O_x^2+O_z^2& \le 1~.
\end{aligned}
\end{equation}
If one now considers these three observables together, it is found that 
\begin{equation}
\Sigma^2+O_x^2+O_z^2 \le 1~.
\end{equation}
This situation is, however, not automatic for three observables with a disk constraint on each pair. For $\ppLL$, we have cases where the domain is the intersection of the three orthogoanl cylinders, which is slightly wider than the unit sphere, see Fig.~\ref{excl:fig:lim3d} (b).

Other triples for which an unit sphere is found as boundary are:
\begin{equation}
\begin{aligned}
P^2+C_x^2+C_z^2 &\le 1~,\\
P^2+O_x^2+O_z^2 &\le 1~,\\
P^2+C_x^2+O_x^2 &\le 1~,\\
P^2+C_z^2+O_z^2 &\le 1~,\\
\Sigma^2+C_x^2+C_z^2 &\le 1~,\\
\Sigma^2+C_x^2+O_x^2 &\le 1~,\\
\Sigma^2+C_z^2+O_z^2 &\le 1~.
\end{aligned}
\end{equation}

\section{Inclusive reactions}\label{sec:incl}
Inequalities can also be derived for the inclusive reactions, when the initial-state particles are polarised, and the spin of  the identified final particle is measured.

For $a+b\to \hbox{anything}$, the helicity $\Delta \sigma_L$ and transversity $\Delta \sigma_T$ asymmetries of the total cross section $\sigma_{\rm tot}$ satisfy
\begin{equation}
\label{incl:eq:6}
|\Delta \sigma_T| \leq \sigma_{\rm tot} + \Delta  \sigma_L/2 ~.
\end{equation}
this giving the domain depicted in Fig.~\ref{incl:fig:sectot}.
\begin{center}
\includegraphics[width=.35\textwidth]{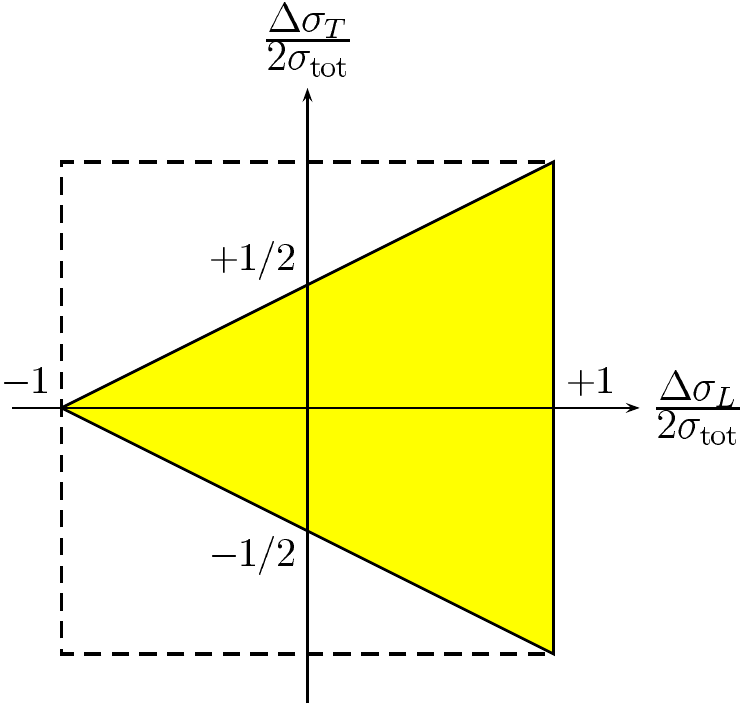}
\end{center}
\figcaption{\label{incl:fig:sectot} Domain allowed for $\Delta \sigma_L$ and $\Delta \sigma _T$}
\begin{center}
\includegraphics[width=.35\textwidth]{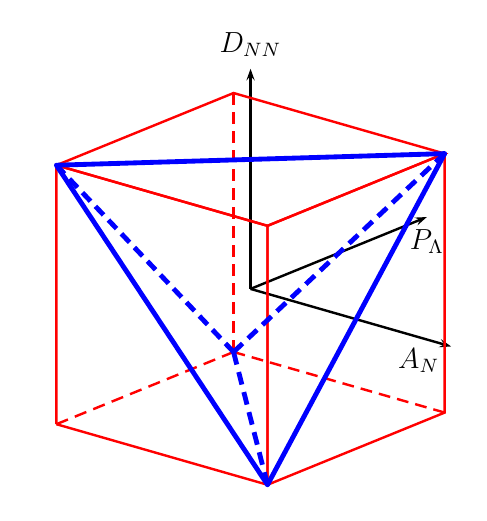}
\end{center}
\figcaption{Allowed domain for $A$, $P_\Lambda$ and $D_{NN}$. 
\label{incl:fig:posit2}}

For $a+b\to c+X$, the simplest observables are the target asymmetry $A_N$, the polarisation $P_\Lambda$ (the notation is inspired from the case where $c=\Lambda$ as in the experiment E-704 at Fermilab, but the result is more general) and depolarisation $D_{NN}$. They are submitted to the same tetrahedron constraint as encountered in exclusive reactions. See Fig.~\ref{incl:fig:posit2}.

Many results deal with the structure functions, the generalised parton distributions and their evolution. In particular, the Soffer inequality 
\begin{equation} q(x)+ \Delta q(x)\ge 2 |\delta q(x)|~,
\end{equation}
which relates the helicity asymmetry $\Delta q$ and the transversity asymmetry $\delta q$ is very similar to (\ref{incl:eq:6}) and gives a  triangular domain identical to Fig.~\ref{incl:fig:sectot}, with the substitution $\Delta \sigma_L/(2\sigma_{\rm tot})\to \Delta q/q$ and $\Delta_T \sigma_T/(2\sigma_{\rm tot})\to \delta q/q$.
\section{Methods}\label{se:meth}
For a given set of amplitudes $a_1$, $a_2$, \dots, the observables are given by bilinear combinations. For instance, in the case of photoproduction
\begin{equation}\label{eq:obs}
\begin{aligned}
 I_0&=|a_1|^2+|a_2|^2+|a_3|^2+|a_4|^2~,\\
I_0 A&=|a_1|^2-|a_2|^2-|a_3|^2+|a_4|^2~\\
I_0 \Sigma&=|a_1|^2+|a_2|^2-|a_3|^2-|a_4|^2~,\\
I_ 0P&=|a_1|^2-|a_2|^2+|a_3|^2-|a_4|^2~\\
 I_0 C_x&=-2 \IM(a_1a_4^*-a_2 a_3^*)~,\\
I_0 C_z&=2 \RE(a_1a_4^*+a_2 a_3^*)~,\\
 I_0 O_x&=-2 \RE(a_1a_4^*-a_2 a_3^*)~,\\
I_0 O_z&=-2 \IM(a_1a_4^*+a_2 a_3^*)~.
\end{aligned}
\end{equation}

In principle, the identities and inequalities among observables are deduced by a mere algebraic manipulation of these expressions. However, such a strategy is not very easy  in practice, and hides a number of features. In particular, the relations are independent of the choice of the set $\{a_i\}$, invariant amplitudes, helicity or transversity amplitudes (though the latter often turns out the easiest to handle).  

Some constraints are just given by common sense and are purely classical. For instance, the sum of squared projections of a polarisation vector cannot exceed unity. As mentioned above, anticommutation relations among operators (of which the observables are the expectation values) 
lead to spherical constraints $X^2+Y^2+\cdots \le 1$. 

There are many symmetries among observables, which are deduced the ones from the others by rotation, crossing symmetry, exchange of particles, etc. Hence only a few basic identities and inequalities need to be established.

The most general method, which is very powerful, consists in writing down the density matrix of a reaction, say $A+B\to C+D+ \cdots$ (the ellipses denotes spinless particles or particles whose spin is not measured), rewritten as $A+B+\overline{C}+\overline{D}\to \cdots$, and to express that this matrix is definite positive. However, this usually involves many observables, and some projection is required to isolate relations among two or three given observables.\cite{Artru:2008cp}\@

For a preliminary investigation, to detect which pairs and triples are submitted to constraints, it is possible to device an empirical  search, which turns out powerful: fictitious amplitudes are randomly generated, and the corresponding amplitudes are plotted the one vs.\ the other. See, e.g., Figs.~\ref{excl:fig:2-3-8-2-4-7}. Once the contours revealed, it can be attempted to demonstrate rigorously the corresponding inequalities.

\section{Outlook}\label{sec:concl}
The progress made on the measurement of spin observables stimulated revisiting the art of the polarisation domain, initiated many years ago by Doncel, Minnaert, Michel, and others. Powerful methods have been developed, in particular to exploit the positivity of the density matrix in any crossed channel. 

Several inequalities  are exploring the \emph{quantum} domain, \ie, go beyond the \emph{classical} inequalities one gets simply by expressing that the outgoing flux is positive for any given configuration of spins.  This means that any hadronic reaction does not escape being a quantum process where a spin state, separable or entangled, undergoes a quantum process and hence is submitted to the rules of transmission of quantum information, in particular these governing the violation of Bell inequalities.

Spin physics has certainly  a rich future, in a variety of energy ranges. In the past, the possibility of measuring the nucleon--nucleon scattering with polarised beam and target enabled one to reconstruct the amplitudes (to an overall phase) and to test the mechanism of nuclear forces. The polarisation measurements on pion--nucleon scattering opened the field of baryon resonances which is the topics of this conference. Thanks to polarisation, some measurements at SLC have challenged those of LEP. Today, the possibility of proton--proton scattering with both beams polarised give unique opportunities to the Brookhaven experiments.  On the contrary, after more than 10 years of data taking at LEAR, ambiguities remain in the low-energy antinucleon--nucleon interaction, due to cuts in the spin-physics programme. 

Dramatic progress has been achieved in the techniques of polarised targets, which are now more stable and more easily implemented into the detectors.  The next challenge is to build new polarised beams. There are serious studies and frequent meetings about the possibility of polarising positron beams. See, e.g., the proceedings of the recent ``Posipol'' workshops.\cite{posipol,MoortgatPick:2006qp}\@ Many mechanisms have been proposed to polarise antiprotons, years ago when it was believed that this physics could be developed at CERN, and more recently\cite{Chattopadhyay:2008zz} in view of performing experiments at FAIR: filtering of polarisation states by a polarised target, transfer of polarisation from a companion beam, etc. There are often controversies, and for the processes based on the nucleon--antinucleon interaction, data on  the relevant observables are lacking. It might be reminded, however, that one can very likely produce \emph{polarised antineutrons} by shooting an antiproton beam on a longitudinally polarised proton target. \cite{Dover:1981pp,Richard:1982au}
\\

\acknowledgments{J.M.R. would like to thank the organisers for this beautiful and stimulating conference.}

\end{multicols}

\vspace{-2mm}
\centerline{\rule{80mm}{0.1pt}}
\vspace{2mm}

\begin{multicols}{2}


\begin{thebibliography}{10}

\bibitem{Bourrely:1980mr}
C.~Bourrely, J.~Soffer, and E.~Leader.
\newblock Polarisation phenomena in hadronic reactions.
\newblock {\em Phys. Rept.}, 59:95--297, 1980.

\bibitem{Leader:2001gr}
E.~Leader.
\newblock Spin in particle physics.
\newblock {\em Camb. Monogr. Part. Phys. Nucl. Phys. Cosmol.}, 15:1, 2001.

\bibitem{Ohlsen:1972xx}
G.G. Ohlsen.
\newblock Polarization transfer and spin correlation experiments in nuclear
  physics.
\newblock {\em Rep. Prog. Phys.}, 35:717--801, 1972.

\bibitem{Artru:2008cp}
X.~Artru, M.~Elchikh, J.-M.~Richard, J.~Soffer, and O.V.~Teryaev.
\newblock {Spin observables and spin structure functions: inequalities and
  dynamics}.
\newblock {\em Phys. Rept.}, 470:1--92, 2009.

\bibitem{Artru:2006xf}
X.~Artru, J.-M.~Richard, and J.~Soffer.
\newblock Positivity constraints on spin observables in exclusive pseudoscalar
  meson photoproduction.
\newblock {\em Phys. Rev.}, C75:024002, 2007.

\bibitem{Lleres:2008em}
A.~Lleres et~al.
\newblock {Measurement of beam-recoil observables $O_x$, $O_z$ and target asymmetry
  for the reaction $\gamma p\to K \Lambda$}.
\newblock {\em Eur. Phys. J.}, A39:149--161, 2009.

\bibitem{posipol}
{\sf http://posipol2009.in2p3.fr, \\
  http://posipol2006.web.cern.ch/Posipol2006/}.

\bibitem{MoortgatPick:2006qp}
Gudrid Moortgat-Pick.
\newblock Physics aspects of polarized $\mathrm{e}^+$ at the linear collider.
\newblock 0700.
\newblock Talk given at the \textsl{POSIPOL2006} Workshop, CERN, April 2006.

\bibitem{Chattopadhyay:2008zz}
S.~Chattopadhyay, D.P.~Barber, N.~Buttimore,
G.~Court,and E.~Steffens (eds.)
\newblock {Polarized antiproton beams - how? Proceedings, International
  Workshop, Daresbury, UK, August 29-31, 2007}.

\bibitem{Dover:1981pp}
C.~B. Dover and J.~M. Richard.
\newblock S\lowercase{PIN OBSERVABLES IN LOW-ENERGY NUCLEON ANTI-NUCLEON
  SCATTERING}.
\newblock {\em Phys. Rev.}, C25:1952--1958, 1982.

\bibitem{Richard:1982au}
J.-M.~Richard.
\newblock S\lowercase{PIN DEPENDENCE} in {\NNb} at low energy.
\newblock Lecture given at Int. Sch. of Physics of Exotic Atoms: Workshop on
  Physics at LEAR with Low Energy Cooled Antiproton, Erice, Sicily, May 9-16,
  1982.

\end{thebibliography}

\end{multicols}
\end{document}